# A dynamically reprogrammable metasurface with self-evolving shape morphing

Yun Bai[a,1], Heling Wang[b,c,d,1,*], Yeguang Xue[b,c,d], Yuxin Pan[a], Jin-Tae Kim[e], Xinchen Ni[e], Tzu-Li Liu[e], Yiyuan Yang[c], Mengdi Han[e,k], Yonggang Huang[b,c,d,e,*], John A. Rogers[b,c,d,e,f,g,h,i,j*], and Xiaoyue Ni[a,e,l,*]

[a]Department of Mechanical Engineering and Materials Science, Duke University, Durham, NC, USA

[b]Department of Civil and Environmental Engineering, Northwestern University, Evanston, IL, USA

[c]Department of Mechanical Engineering, Northwestern University, Evanston, IL, USA

[d]Department of Materials Science and Engineering, Northwestern University, Evanston, IL, USA

[e]Simpson Querrey Institute for Bioelectronics, Northwestern University, Evanston, IL, USA

[f]Department of Biomedical Engineering, Northwestern University, Evanston, IL, USA

[g]Department of Chemistry, Northwestern University, Evanston, IL, USA

[h]Department of Neurological Surgery, Feinberg School of Medicine, Northwestern University, Chicago, IL, USA

[i]Department of Electrical and Computer Engineering, Northwestern University, Evanston, IL, USA

[j]Department of Computer Science, Northwestern University, Evanston, IL, USA

[k]Department of Biomedical Engineering, College of Future Technology, Peking University, Beijing, China

[l]Department of Biostatistics and Bioinformatics, Duke University, NC, USA

[1]These authors contribute equally

[*]Correspondence should be sent to: helingwang1@gmail.com (H.W.) or y-huang@northwestern.edu (Y.H.) or jrogers@northwestern.edu (J.A.R.) or xiaoyue.ni@duke.edu (X.N.)

**Abstract**

Dynamic shape-morphing soft materials systems are ubiquitous in living organisms; they are also of rapidly increasing relevance to emerging technologies in soft machines[1–4], flexible electronics[5–7], and smart medicines[8,9]. Soft matter equipped with responsive components can switch between designed shapes or structures, but cannot support the types of dynamic morphing capabilities needed to reproduce natural, continuous processes of interest for many applications[10–27]. Challenges lie in the development of schemes to reprogram target shapes post fabrication, especially when complexities associated with the operating physics and disturbances from the environment can prohibit the use of deterministic theoretical models to guide inverse design and control strategies[3,28–32]. Here, we present a mechanical metasurface constructed from a matrix of filamentary metal traces, driven by reprogrammable, distributed Lorentz forces that follow from passage of electrical currents in the presence of a static magnetic field. The resulting system demonstrates complex, dynamic morphing capabilities with response times within 0.1 s. Implementing an in-situ stereo-imaging feedback strategy with a digitally controlled actuation scheme guided by an optimization algorithm, yields surfaces that can self-evolve into a wide range of 3-dimensional (3D) target shapes with high precision, including an ability to morph against extrinsic or intrinsic perturbations. These concepts support a data-driven approach to the design of dynamic, soft matter, with many unique characteristics.

**Results**

Soft matter that can dynamically reconfigure their shapes upon interactions with environment or perceptions of information is thriving[33]. Pioneering studies rely on an exploitation of responsive materials or material configurations to create active structures that shift their shapes in response to external stimuli[34–38]. Smart materials (*e.g.,* liquid crystal elastomers[11,13–17,39,40], shape memory polymers[41], hydrogels[10,12,24], and others[25]) and multimaterial structures[11,26] enable large structural deformation but face challenges in implementing fast control to refined structures. The design of shape-morphing process usually requires prerequisite modeling effort to be programmed into the fabrication process, and is therefore hard to reprogram on-the-fly (*e.g.,* 3D printing[11,27], magnetization[19,42], laser or wafer-jet cutting[29,30,43], mechanical buckling[28]). The desire to swiftly shift shapes among large number of configurations post fabrication invites the investigations on programmable stimulus (*e.g.,* temperature[13, 44], magnetic field[20], electric current[22,23]). However, limitations remain in the accessible design space and the real-time inverse design because of the challenges in establishing analytical solutions or barriers in high computational costs due to the complexity arising from nonlinearity or high dimensionality. Also, existing computer-aided methods usually leave the inclusion of imperfections, damages, or the coupling between the system with the unforeseen environment. Incorporating instant feedback is necessary for the morphing process to see the deployment scheme to precisely account for specific, multifunctional, or time-varying requirements[45]. The time constraints and the complexity in actuation, feedback, or modeling all contribute to a prolonged programming cycle that limits the possible shapes or shape responses to remain discrete and quasi-static.

    Here, we demonstrate a dynamically reprogrammable mechanical metasurface with a closed-loop 3-dimensional (3D) shape control, based on a digital, fast, and precise Lorentz force actuation scheme. The metasurface takes the form of interconnected,

serpentine-shaped beams that consist of a thin conductive layer of gold (Au, thickness $h_{Au}$ = 300 nm) encapsulated by polyimide (PI, thickness $h_{PI}$ = 7.5 μm, width $b_{PI}$ = 160 μm) (see Methods section 'Sample fabrication', Supplementary Note S1, and Supplementary Fig. 1 for details). The intersections of the beams form an N×M mesh as shown in Fig. 1a (N = M = 4, sample size L = W = 18.0 mm). A tailored serpentine design ensures sufficiently large, fast, and reversible out-of-plane deformation ($u/L$~30%; response time <0.07 s) of the serpentine beam, driven by a modest electric current (I < 27.5 mA) in a magnetic field **B** (magnitude B = 0.224 ± 0.016 T) (see Supplementary Notes S2–5 and Supplementary Figs. 2–7 for details). Fig. 1b shows that independent voltages (**V**={$V_j$}) of size 2(N+M) applied to the peripheral ports, controls the distribution of current density (**J**) in the conductive network (see Methods section 'Digital control' and Supplementary Fig. 8 for details) and therefore the Lorentz force $\boldsymbol{F}_{EM} = \boldsymbol{J} \times \boldsymbol{B}$. The spatially distributed actuation $\boldsymbol{F}_{EM}(\boldsymbol{J})$ controls the local, out-of-plane (Z-direction) deformations (**u** = {$u_i$}, where $u_i$ is the displacement of the $i^{th}$ node) of the sample in a magnetic field **B** aligned with its diagonal, enabling a large set of accessible 3D shapes from the same precursor structure. The unusual structure and material design further enables the system to adopt an approximate, linearized model, such that the nodal displacement response to the input voltages follows,

$$u_i = \sum_{j=1}^{2(N+M)} C_{ij}V_j, \text{ for } i = 1, \dots N \times M, \quad (1)$$

where the coupling matrix $\boldsymbol{C} = \{C_{ij}\}$ fully characterizes the electro-magneto-mechanical system. Fig. 1c shows the finite element analysis (FEA) and the experimental characterization of the coupling coefficients $C_{ij}$ for representative nodes in the actuation range of 0–4 V for the 4×4 sample in the magnetic setup. Linear regression on the FEA results predicts $\boldsymbol{C}$. The analytical model and the FEA studies provide a scaling law of the coefficients as $C_{ij} \sim (BLH^2 b_{Au} h_{Au})/(E_{PI} b_{PI} h_{PI}^3 \rho_{Au})$ (H–serpentine beam width, $E_{PI}$–PI

Young's modulus, $\rho_{Au}$–Au electrical conductivity; see Supplementary Notes S3.2, 3.3, Supplementary Figs. 9, 10 for details). Following this linear approximation, a model-driven approach attempts to zero the errors, $e_i(\boldsymbol{V}) = (u_i(\boldsymbol{V}) - u_i^*)/L$ (difference between the output deformation, $u_i(\boldsymbol{V})$, from the target, $u_i^*$, normalized by system size $L$), to optimize the voltages for the precursor surfaces to deform to target implicit shapes. Specifically, a gradient-descent based algorithm iterates over $\boldsymbol{V}$ to minimize a loss function, $f(\boldsymbol{V}) = \sum_i e_i^2(\boldsymbol{V})$ with a maximum-current constraint (see Methods section 'Optimization algorithm' in Methods and Supplementary Note S6 for details). The linearized model-driven approach yields a prediction for $\boldsymbol{V}$ within ~10 ms. The same approach driven by numerical methods (e.g., FEA) without linearization is not possible due to unaffordable computational costs (~10 days using a workstation with 40-core, 2.4 GHz CPU, and 64 GB memory). Fig. 1d shows FEA and experimental results of an inverse-designed, continuous shape morphing of a 4×4 and an 8×8 sample ($L = W = 22.4$ mm). The process consists of four phases: growing up, moving around, splitting and oscillating, with a prescribed control of the instantaneous velocity and acceleration of the dynamics (Supplementary Video 1, Supplementary Note S7 and Supplementary Figs. 11–14).

In addition to the abstract, implicit shapes, the reprogrammable metasurface demonstrates an ability to reproduce dynamic processes in nature that involve a temporal series of complex shapes, provided with the inversely designed current distributions. Fig. 2a shows an array of 8 serpentine beams ($L = 10.4$ mm, $W = 20.6$ mm, Supplementary Note S8 and Supplementary Fig. 15) morphing into the 2-dimensional profile of a droplet dripping from a nozzle (Supplementary Video 2 and Supplementary Fig. 16). Shapes I–III describe the growing of a pendant drop to its critical volume. Shapes IV–V capture the following pinch-off process. Fig. 2b and c present the 4×4 and 8×8 samples simulating the 3D surface of a droplet hitting a rigid surface in five stages: falling on the surface, spreading out, bouncing back, vibrating and stabilizing (Supplementary Video 3 and

Supplementary Figs. 17–20). Numerical analysis further illustrates that the mesh structure can morph into an extensive set of target shapes (see Supplementary Notes S7, S9, S10, and Supplementary Figs. 21–28).

The linearized model-driven approach accomplishes the inverse design when a modest error from the nonlinearity is tolerable. Extending the model-driven approach to include nonlinearity is challenging due to large computational expenses (Supplementary Note S11) or difficulties in establishing analytical solutions. The open-loop model-based inverse design has constraints in the design space and cannot account for non-ideal factors such as environmental changes or defects in the sample. The existing limitations motivate the development of sensing feedback for a closed-loop self-evolving approach.

Fig. 3a illustrates an experiment-driven process in comparison with the linearized model-driven process. While the model-driven route relies on the presumption of a linear and stationary model, the experimental method takes the in-situ measurement of the system output and feed the difference between the current states and the target states for actuation regulation. In this work, a custom-built stereo-imaging setup using two webcams (ELP, MI5100) enables a 3D reconstruction of the nodal displacement at a rate of 30 frames per second (fps), with a displacement resolution of ~0.006 mm and a measurement uncertainty of ±0.055 mm (see details of 3D imaging in Methods, Supplementary Note S12 and Supplementary Fig. 29). After each update of the actuation (***V***), the real-time imaging provides an in-situ nodal displacement error analysis. An optimization algorithm, same as the one used in the model-driven approach but wrapping the 3D imaging process, performs the experimental iterations over $\boldsymbol{V}$ to minimize $f(\boldsymbol{V})$. For a 4×4 sample morphing into a representative target shape ($f(\boldsymbol{V} = \boldsymbol{0})$ = 0.05–0.35), the optimization process takes 5–15 iterations (Supplementary Figs. 30). Each feedback-control cycle in the current setup takes ~0.25 s due mainly to the time overhead from imaging processing algorithm but is ultimately limited to the mechanical response time (<0.1 s) (Supplementary Note S6 and Supplementary Table 1). A hybrid method, taking a model-driven prediction as the initial input, reduces the number of iterations to ~3 at the cost of a preceding modeling effort. The dominant sources of errors are discreteness in input voltages and uncertainties associated with 3D imaging (Supplementary Note S12 and Supplementary Fig. 31). The experiment-driven process opens opportunities for the metasurface to self-evolve to target shapes without any pre-knowledge of the system (Supplementary Video 4). Fig. 3b provides a quantitative comparison between the model-driven and experimental-driven morphing results from the same 4×4 precursor, targeting representative implicit shapes (Supplementary Note S7, Supplementary Figs. 32–35, and Supplementary Video 5). The resulting errors from the model-driven approach follows a wide (over ±5%), mostly skewed distribution (considering 441 points from the interpolated 3D surface; Supplementary Note S12). The experiment-driven approach, accounting for the subtle nonlinear deviation, yields a relatively narrow (±2%), symmetric error distribution.

The experiment-driven process works as a physical simulation to accommodate pronounced nonlinearity without a significant increase in the computational cost. Fig. 4a introduces a 2×2 sample ($L = W = 25.0$ mm) consisting of serpentine beams with the relative arc length reduced morphing into the same target shape in Fig. 3b. Centered in the same magnetic setup, the sample exhibits an amplified non-linear mechanical behavior in response to input voltages (Supplementary Note S13 and Supplementary Fig. 36). The model-driven approach based on the linear-system assumption results in an absolute maximum error of ~8%. The experimental-driven approach achieves more accurate morphing result in ~20 iterations with absolute errors below 1%.

Guided by the experiment-driven process, the metasurface can also self-adjust to morph against unknown perturbations. Fig. 4b–d shows three representative cases in which a 4×4 sample morphs with perturbed magnetic field, external mechanical load, and intrinsic damage, respectively. In all cases, the model-driven approach following the original inverse design results in absolute maximum errors of ~8–10%. In comparison, the experiment-driven approach adapts the shape to reach the target with absolute errors below ~3% that is comparable with that of an intact sample (~2%) (Supplementary Video 6). The significantly boosted accuracy level demonstrates a 'self-sustained' morphing ability enabled by the experiment-driven process.

The adaptive, self-evolving metasurface platform delivers a semi-real-time morphing scheme to learn the continuously evolving surface of a real object in-time. In this experiment, a duplicated stereo-imaging setup measures the displacement of a 4×4 array of markers (with inter-spacing $a_0 = 15$ mm) on the palm (Supplementary Fig. 37a). The optimization acts directly to minimize the displacement difference between the 16 markers and their corresponding nodes of a 4×4 sample. Given continuity, the gradient-descent process takes the last morphing result as the initial state for the next morphing task. This differential method (with the target descent $\Delta f(V)$ ~0.08) requires only ≤3 iterations (~20 s) to reach the optimum. Fig. 5a shows representative frames from a video recording of hand making eight gestures with different fingers moving (see Supplementary Fig. 37b, c and Supplementary Video 7 for complete results of all gestures). All morphing results agree with the target with absolute errors below 2%.

In addition to self-evolving to optimize shapes, the metasurface can self-evolve to optimize functions. Setting multiple target functions drives the optimization towards emergent multifunctionality, with an ability to decouple naturally coupled functions. Fig. 5c illustrates a scheme where a 3×3 sample ($L = W = 14.8$ mm) with 9 reflective gold patches at the nodes, attempts to perform an optical and a structural function: I) reflect and overlap two laser beams (red, green) with different incident angles ([$\theta_{Xr}$ $\theta_{Zr}$], [$\theta_{Xg}$ $\theta_{Zg}$]) on a receiving screen (Supplementary Fig. 38a) and II) achieve the target displacement of its central node. The optimization takes a hybrid strategy combining the model-driven and experiment-driven processes (Supplementary Note S14). While the voltages control the reflected beam paths, a top camera provides an imaging feedback of the distances between the beam spots on the screen. The model-driven process predicts the difference between the central nodal displacement and the target. The total loss takes a linear combination of the two errors (Supplementary Note S14, Supplementary Fig. 38b). Fig. 5d shows the self-evolving results of three optical configurations with distinctive incident beam angles. Fig. 5e shows that the metasurface can morph to overlap the laser spots on the receiving screen with a range of possible shapes (Supplementary Fig. 39a). By enforcing both functions, the sample overlaps the spots and settles its central node to a target displacement. A post analysis via ex-situ 3D imaging validates that the final experimental central nodal displacement reaches the target within an error of ±2% (Supplementary Fig. 39b and Supplementary Video 8).

The work presents a reprogrammable metasurface that can precisely and rapidly morph into a wide range of target shapes and dynamic shape processes. The highly integrable digital-physical interfaces incorporating actuation, sensing, and feedback allow for an in-loop optimization process to guide the metasurface to self-evolve to target shapes, without prior knowledge of physics, or with a model-driven prediction to expedite the evolving process. Such scheme enables an autonomous materials platform to promptly change structures, actively explore the design space, and responsively reconfigure functionalities towards unprecedented performance and efficiency. The experiment-driven shape shifting capability addresses existing theoretical and computational challenges in complex, nonlinear systems, bringing new opportunities to physical simulations for a real-time, data-driven inverse design process. Compared to existing shape morphing or structural reconfiguration methods that are often slow in programming cycles[18], exclusive to responsive polymeric materials or complex structures[11,13], or require customized fabrication procedures that are difficult to scale[19,21], our Lorentz-force strategy enables extremely fast, precise, and digitally reprogrammable soft matter that is compatible with the typical materials, structures, and thin-film fabrication techniques used in the existing flexible electronics framework. The usage of conventional conductive materials and the potential scalability of the platform promise a wide, versatile application scenario in wearable techniques, soft robotics, and advanced materials. Many possibilities exist to improve this system. For example, incorporating a mechanically-locking mechanism (*e.g.*, applying phase transition materials[21,46] or jamming configuration[47] could hold the morphed shapes without actuation. The current modular platform demonstration invites future work to embed sensing, computing, and communicating functions directly into the materials for higher levels of integration. Employing advanced data-driven techniques in the loop, (*e.g.*, Bayesian optimization[48], Deep Learning[49], Reinforcement Learning[50]) will bring the self-evolving morphing ability of artificial matter to a level closer to or even beyond their natural counterpart, paving a way for new classes of intelligent materials that adopt spatiotemporally controlled shapes and structures for advanced on-demand functionalities.

**Methods**

**Sample fabrication.** The fabrication process (Supplementary Fig. 1) began with spin coating a thin layer of PI (HD Microsystems PI2545, ~3.75 µm in thickness) on a silicon wafer with poly(methyl methacrylate) (PMMA; Microresist 495 A5, ~80 nm in thickness) as the sacrificial layer. Subsequent lift-off processes patterned the metal electrodes and serpentine connections (Ti/Au, 10 nm/300 nm in thickness). Spin coating another layer of PI (HD Microsystems PI2545, ~3.75 µm in thickness) covered the metal pattern. Photolithography and oxygen plasma etching of PI defined the outline of sample. Undercutting the bottom layer of PMMA allowed transfer of the sample to a water-soluble polyvinyl alcohol (PVA) tape (3M) from the silicon wafer.

**Digital control.** The digital control system used (i) pulse-width modulation (PWM) drivers (PCA9685, 16-channel, 12-bit), (ii) voltage amplifier circuits (MOSFET, IRF510N, Infineon Tech), and (iii) a single-board computer (Raspberry Pi 4) remotely connected to an external computer (Intel NUC, Intel Core i7-8559U CPU@2.70GHz). The external computer ran the optimization algorithm and sent the updated values of voltage profile wirelessly to the single-board computer through Python Socket network programming.

The PWM driver received the actuation signals from the single-board computer. Each PWM channel, operated at a frequency of 1,000 Hz, generated an independent voltage in the range of 0-6 V with 12-bit (~0.0015 V) resolution. The single stage MOSFET provided a reversely linear amplification to the PWM output with a gate voltage, $V_{gs(th)}$ = 4 V, and an external power supply, $V_{ex}$ = 6 V (Supplementary Fig. 8).

**Optimization algorithm.** Sequential Least Squares Programming (SLSQP) with 3-point method (SciPy-Python *optimize.minimize* function) computed the Jacobian matrix in the loop to minimize the loss function $f(V)$. The model-driven approach adopted the same optimization algorithm, with $f(V)$ evaluated by Eq. (1) and a maximum of ~10,000 iterations. For the experiment-driven approach, a maximum final loss value of $0.005f(V = 0)$ and a maximum of 15 iterations set the stopping criteria for the optimization process. Each iteration required 2(*N+M*+1*)* function evaluations for an *N*×*M* sample (Supplementary Note S6).

**3D imaging.** The multi-view stereo-imaging platform consisted of two cameras (Webcams, ELP, 3840×2160-pixel resolution, 30 fps) connected to the external computer taking top-view images of the sample from symmetric angles (Supplementary Fig. 29a). A calibration algorithm (OpenCV-Python *calibrateCamera* function) applied to a collection of images of a checkerboard (custom-made, 7×8 squares, 2 mm × 2 mm per square) returned camera matrix, distortion coefficients, rotation and translation vectors to correct for the lens distortion of the images (OpenCV-Python *undistort* function). The nodes of the mesh samples provided distinguishable geometry for image registration. A template matching algorithm (OpenCV-Python *matchTemplate* function) returned the locations of the nodes in the images from the two cameras. A perspective projection algorithm (OpenCV-Python *reprojectImageTo3D* function) transformed the disparity map to the nodal heights in a unit of pixels ($u_\mathrm{p}$). An additional side camera provided ground-truth measurement of the displacement ($u_\mathrm{m}$) of the discernible nodes and provided a linear-model prediction of the 3D-recontructed nodal displacement ($u(u_\mathrm{p})$) (Supplementary Fig. 29b, c and Supplementary Note S12).

## Data availability

All data are contained within the manuscript. Raw data are available from the corresponding authors upon reasonable request.

## Code availability

The codes that support the findings of this study are available from the corresponding authors upon reasonable request

## Acknowledgements

Y.B., Y.P., and Xiaoyue Ni acknowledge funding support from the Pratt School of Engineering and School of Medicine at Duke University. Y.H. acknowledges support from the NSF (CMMI 16-35443).

## Author contributions

Y.B., H.W., Y.H., J.A.R., and Xiaoyue Ni conceived the idea and designed the research. Y.B. and Y.Y. fabricated the samples. Y.B., Y.X., Y.P., J.K., Xinchen Ni, T.L., M.H., and Xiaoyue Ni performed the experiments. H.W. and Y.H. performed the finite-element modeling and theoretical study. Y.B. and Xiaoyue Ni, analyzed the experimental data. Y.B., H.W., Y.H., J.A.R., and Xiaoyue Ni wrote the manuscript, with input from all co-authors.

**Competing interests** The authors declare no competing interests.

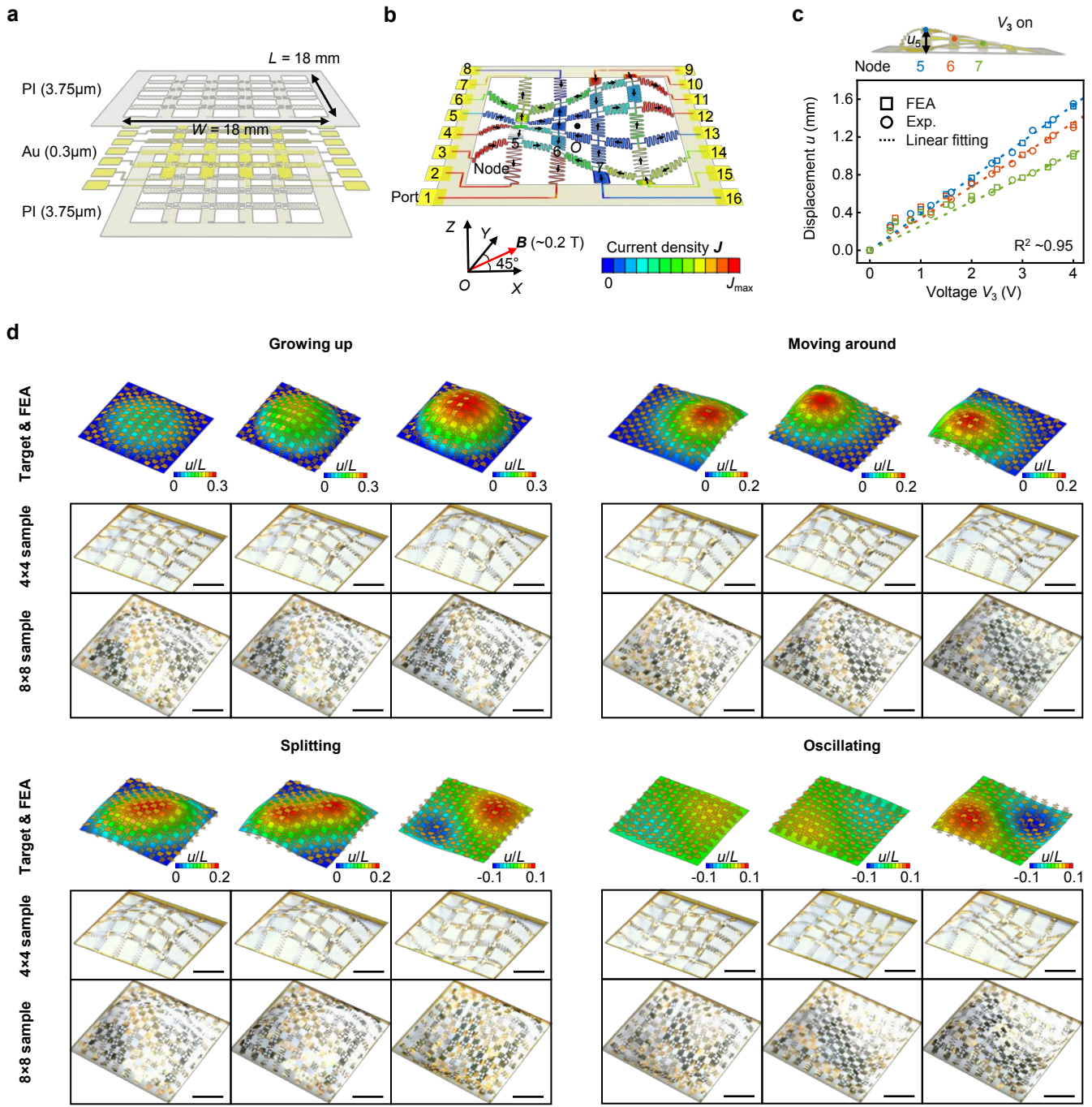

**Fig. 1 | Mechanical metasurfaces driven by reprogrammable electromagnetic actuation. a**, Schematic illustration (exploded view) of a representative square mesh sample constructed from the serpentine beams consisting of thin polyimide (PI) and gold (Au) layers. **b**, Schematic illustration of a 4×4 sample (column and row serpentine length $L_{N/M}$ = 2.5 mm) placed in a magnetic field (in-plane with the sample in a diagonal direction). Port voltages define the current density distribution ($J$) in the sample and hence control the local Lorentz force actuation. **c**, Finite element analysis (FEA) provide a linear-model approximation of the nodal displacement in response to the input voltages for the 4×4 sample. Experimental characterization using a side camera agrees with the FEA prediction. **d**, FEA and experimental results of a 4×4 and 8×8 sample morphing into four target implicit shape shifting processes with control on instantaneous velocity and acceleration of the dynamics. Scale bars, 5 mm.

Figure 1

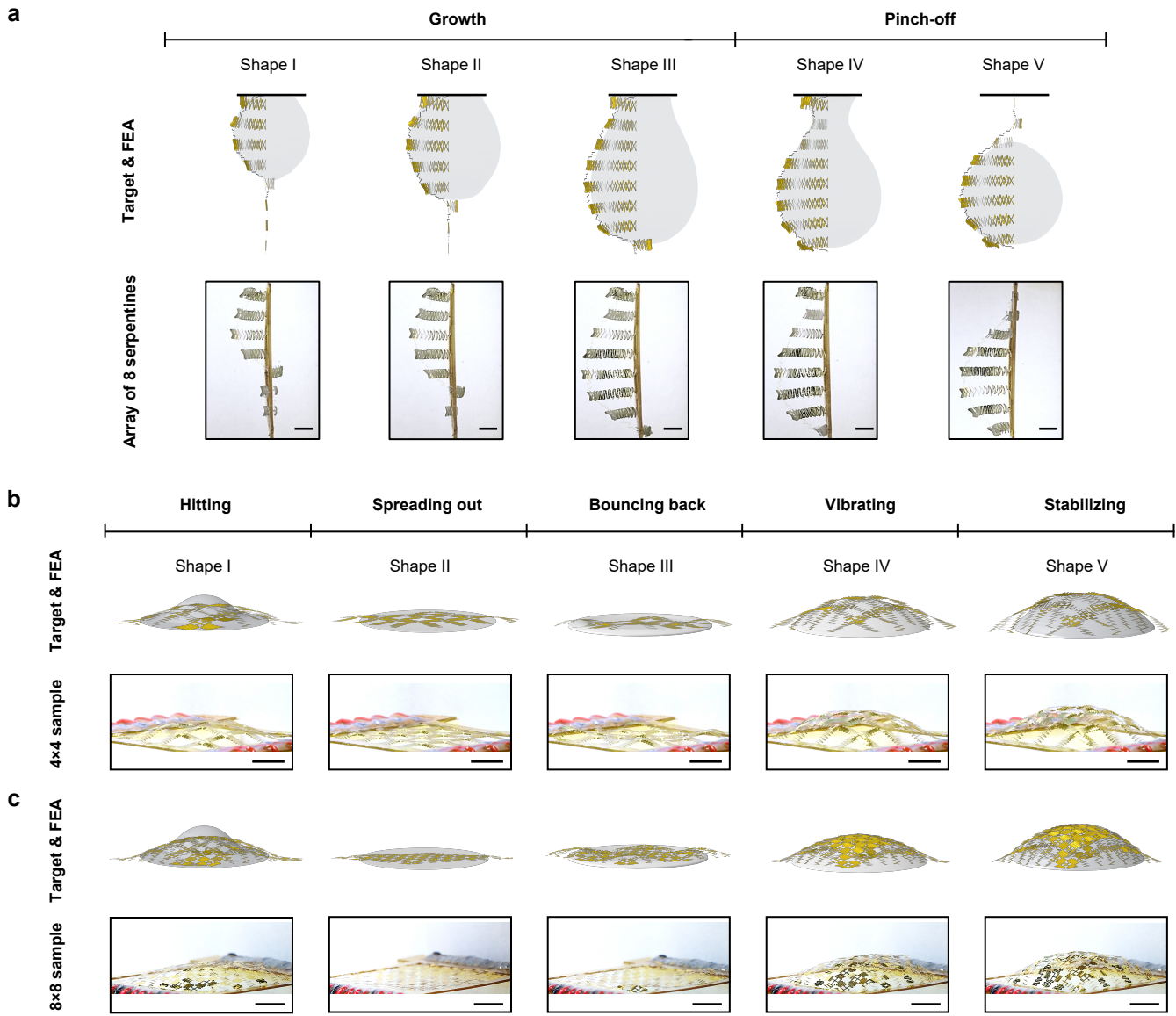

**Fig. 2 | Model-driven inverse design of the metasurfaces for dynamic, complex shape morphing. a**, FEA and experimental results of an array of 8 serpentine beams morphing into the growth and pinch-off of a droplet dripping from a nozzle. **b**, **c**, FEA and experimental results of a 4×4 (b) and an 8×8 (c) sample reproducing the dynamic process of a droplet hitting a solid surface, spreading out, bouncing back, vibrating, and stabilizing.

Figure 2

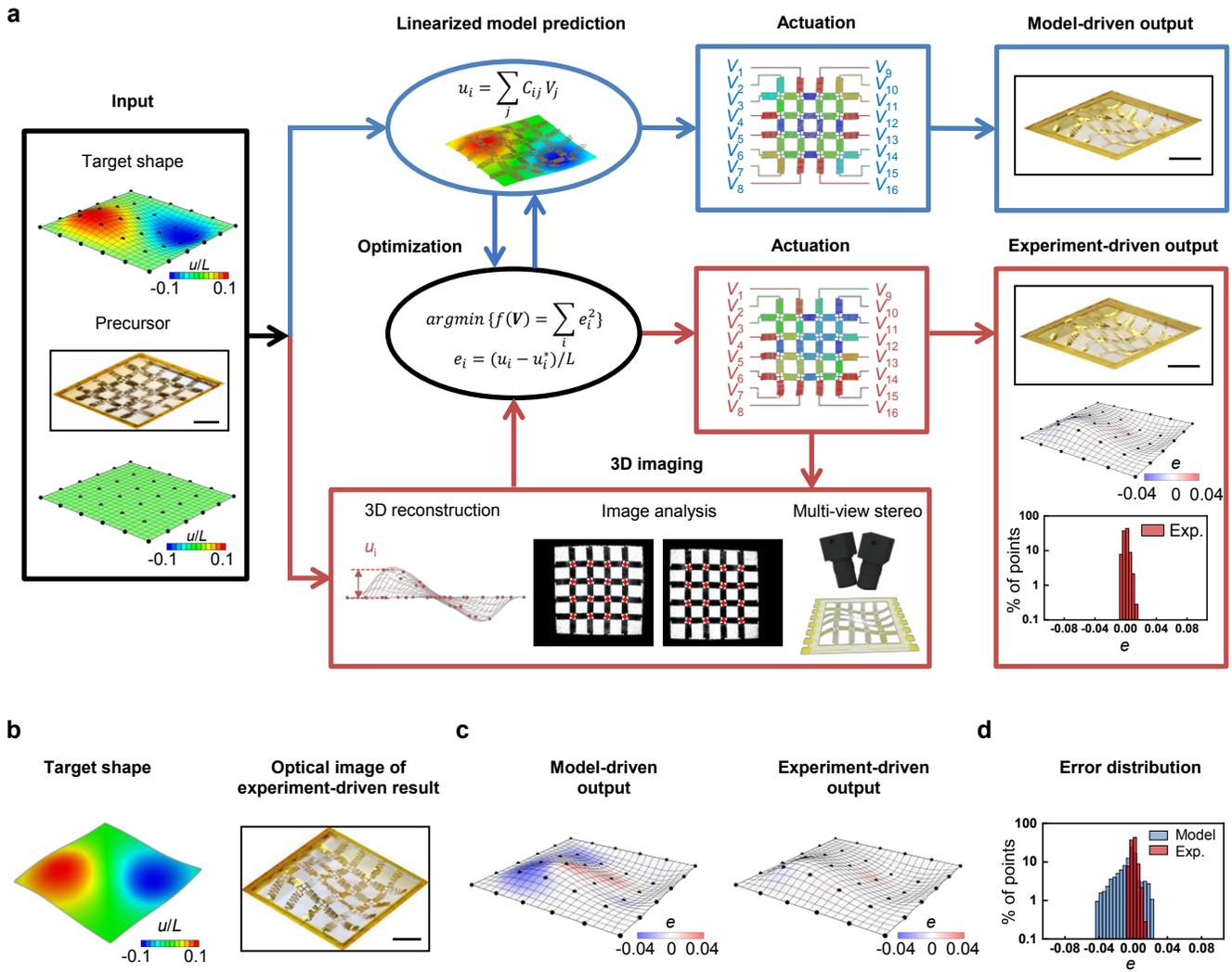

**Fig. 3 | Experiment-driven self-evolving process in comparison with the model-driven approach. a**, Flow diagram of the model-driven inverse design approach (top, blue) and an experiment-driven self-evolving process enabled by an in-situ 3D-imaging feedback and a gradient-descent based optimization algorithm (bottom, red). **b**, Target implicit shapes and optical images of the experiment-driven morphing results of a 4×4 sample. **c**, 3D reconstructed surfaces overlaid with contour plots of the minimized errors ($e$) and **d**, histograms of the minimized errors for model-driven and experiment-driven outputs. Scale bars, 5 mm.

Figure 3

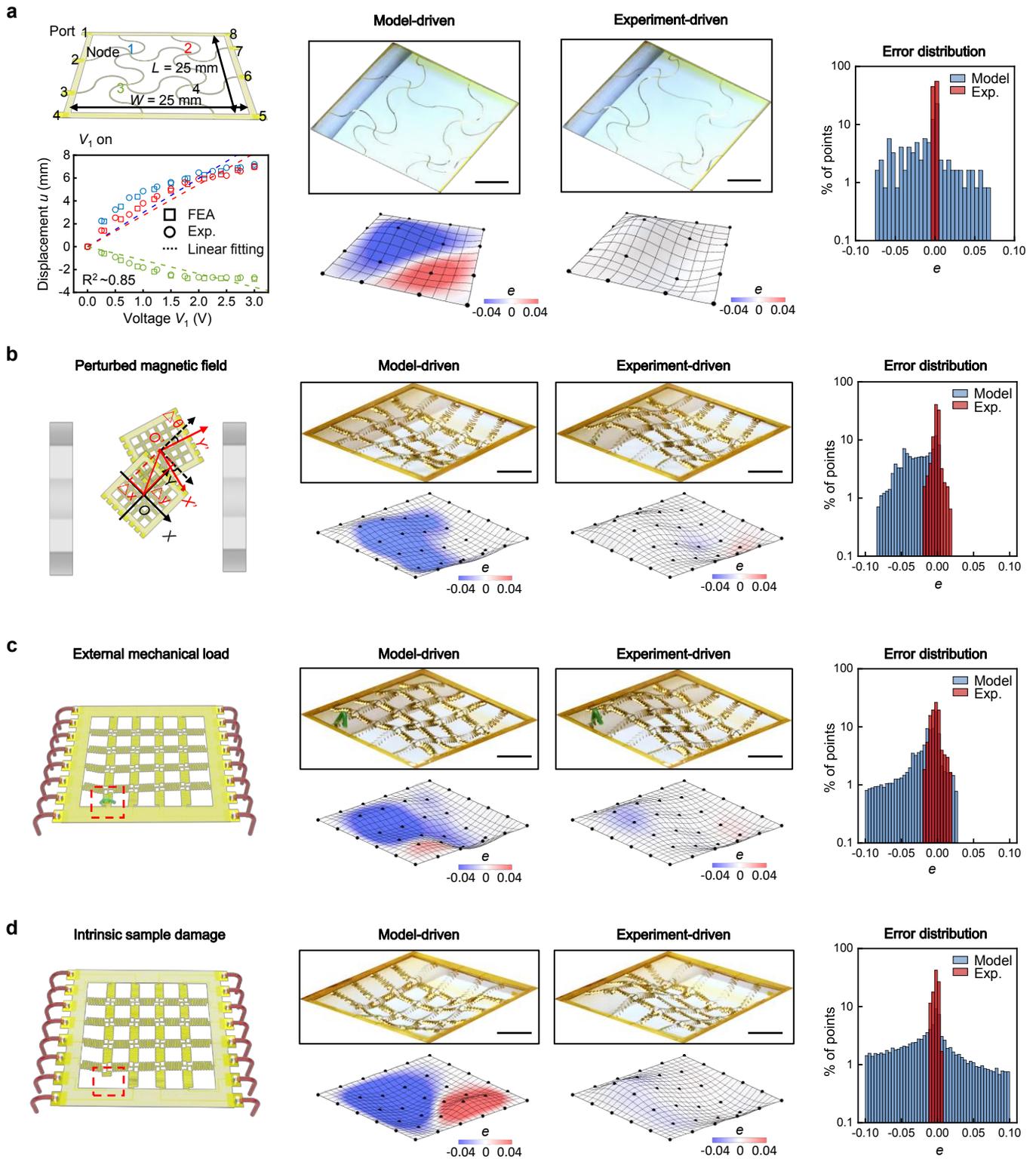

**Fig. 4 | Self-evolving shape morphing against extrinsic or intrinsic perturbations. a**, **b**, **c**, **d**, Experimental results of a 2×2 sample (a) and a 4×4 sample (b-d) morphing into the same target shape (Fig. 3b) via model-driven and experiment-driven processes with a modified serpentine design that amplifies the non-linearity of the voltage-driven deformation (a), and an introduction of an extrinsic magnetic perturbation by displacing the sample from the original, centered position (Δ$x$ = 8 mm, Δ$y$ = 12 mm, Δ$θ$ = 15°) (b), an extrinsic mechanical perturbation by applying an external mechanical load (~0.5 g) on a serpentine beam (c), and an intrinsic damage by cutting one beam open, causing substantial changes in both mechanical and electrical conductivity of the sample (d). Left: schematic illustration of the experimental configuration. Middle: optical images and 3D reconstructed surface superimposed with error map (middle). Right: histogram plots of error $e$ (right). Scale bars, 5 mm.

Figure 4

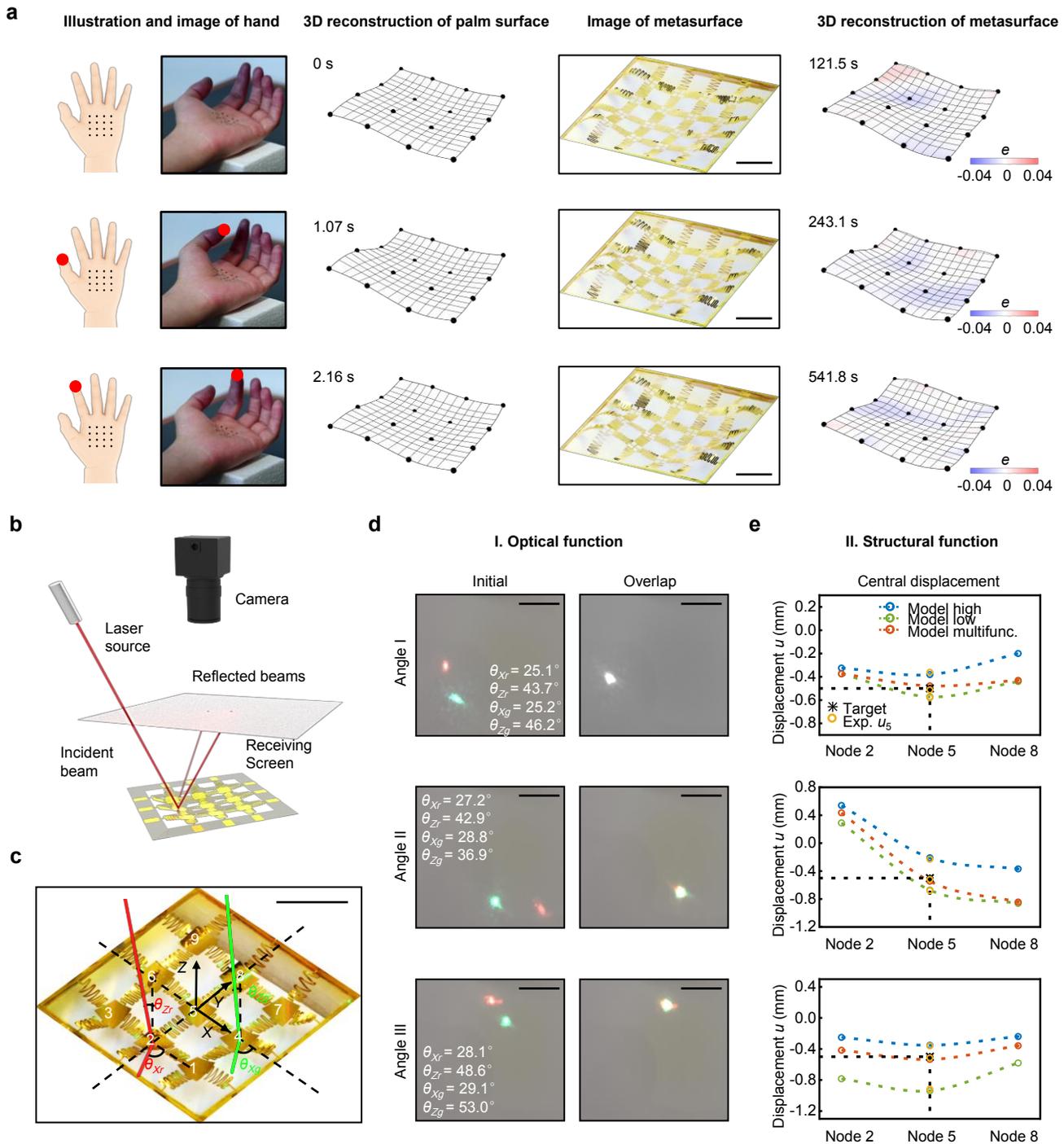

**Fig. 5 | Self-evolving shape morphing toward semi-real-time shape learning and multifunctionality. a**, Morphing results of representative frames from a recording of hand making eight gestures with different fingers moving. **b**, Schematic illustration of a 3×3 sample with gold patches mounted on the nodes reflecting a laser beam from an incident angle. A top-positioned camera monitors the laser spot projected on a paper screen. **c**, A representative optical image of a 3x3 sample ($L_{N/M}$ = 2 mm) with 9 reflective gold patches (Au, 2 mm × 2 mm in size, 300 nm in thickness) at the nodes self-evolving via a hybrid experiment-driven and model-driven process to perform two functions: I) reflect and overlap two laser beams (red, green) with different incident angles ([$\theta_{Xr}$ $\theta_{Zr}$], [$\theta_{Xg}$ $\theta_{Zg}$]) and II) achieve the target displacement (-0.5 mm) of its central node ($u_5$). **d**, Imaging of the screen from the camera provides an experimental feedback of the distance between the two laser spots. **e**, Model predictions of the displacement profile of the sample (cross-sectional view, dashed line in (d)) when overlapping the laser spots with the highest-possible (blue), lowest-possible (green), and optimized (orange) central positions. Ex-situ stereo imaging provides 3D reconstructed measurement of the optimized deformation (black) that validates the in-situ model predictions. Scale bars, 5 mm.

Figure 5